\documentclass[letter,twocolumn]{jpsj3} 
\usepackage{color}

\title{
Magnetization Process 
of the $S=1/2$ Heisenberg Antiferromagnet 
on 
the Cairo Pentagon Lattice
}
\catcode`\@=11
\def\simle{\mathrel{\mathpalette\@versim<}}   
\def\simge{\mathrel{\mathpalette\@versim>}}   
\def\@versim#1#2{\lower2.5pt\vbox{\baselineskip0pt \lineskip-.5pt
   \ialign{$\m@th#1\hfil##\hfil$\crcr#2\crcr\sim\crcr}}}
\catcode`\@=12

\author{Hiroki Nakano$^{1}$
\thanks{E-mail address: hnakano@sci.u-hyogo.ac.jp}, 
Makoto Isoda$^{2}$,  
\thanks{E-mail address: misoda@ed.kagawa-u.ac.jp}
and 
T\^oru Sakai$^{1,3}$
\thanks{E-mail address: sakai@spring8.or.jp}, 
}

\inst{$^{1}$Graduate School of Material Science, 
University of Hyogo,
Kouto 3-2-1, Kamigori, Ako-gun, Hyogo 678-1297, Japan \\
$^{2}$
Department of Physics, Faculty of Education, Kagawa University
Saiwai-cho 1-1, Takamatsu 760-8522, Japan \\
$^{3}$
Japan Atomic Energy Agency, SPring-8, 
Kouto 1-1-1, Sayo, Hyogo 679-5148, Japan 
}

\recdate{\today}

\abst{
We study 
the $S=1/2$ Heisenberg antiferromagnet on 
the Cairo pentagon lattice 
by the numerical-diagonalization method.  
We tune the ratio of two antiferromagnetic interactions 
coming from two kinds of inequivalent sites in this lattice,  
examining the magnetization process of the antiferromagnet; 
particular attention is given to 
one-third of the height of the saturation. 
We find that quantum phase transition 
occurs at a specific ratio 
and that 
a magnetization plateau appears 
in the vicinity of the transition.  
The plateau is accompanied by a magnetization jump 
on one side among the edges due to the spin-flop phenomenon. 
Which side the jump appears depends 
on the ratio. 
}

\begin{document}
\maketitle


Frustration often plays an essential role 
in the behaviors of various magnetic materials, 
where exotic phenomena are induced owing to such frustration. 
Frustration occurs in magnets 
when the system includes a particular geometry of the structure. 
The systems of the kagome-lattice antiferromagnet, 
triangular-lattice antiferromagnet, 
and pyrochlore-lattice antiferromagnet 
which are typical frustrated magnets, 
are composed of local triangles 
creating a strong frustration. 
In spite of the fact that 
extensive studies have been carried out 
to understand these systems well, 
there remain unclear behaviors; 
investigations continue on from various viewpoints 
and by various methods until now. 

\begin{figure}[htb]
\begin{center}
\includegraphics[width=7cm]{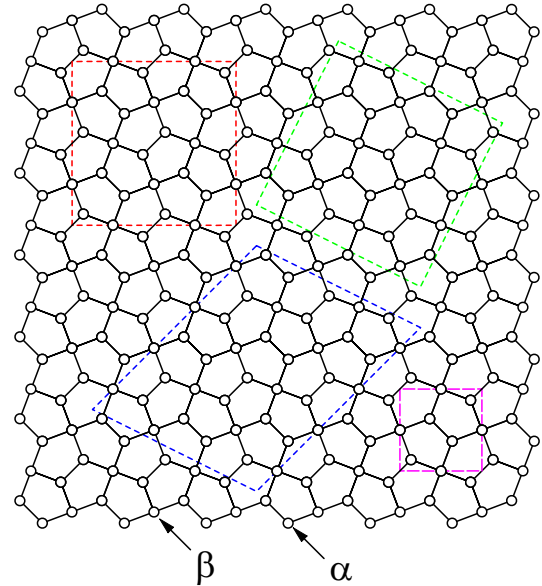}
\end{center}
\caption{(Color) 
Cairo pentagon lattice and its finite-size clusters. 
The small violet square composed of long-dashed lines 
illustrates a unit cell of the Cairo pentagon lattice. 
$\alpha$ and $\beta$ denote a site with 
the coordination numbers $z=3$ and 
$z=4$, respectively. 
Red square, green tilted square, 
and blue parallelogram composed of dotted lines 
denote a finite-size cluster 
for $N_{\rm s}=24$, 30, and 36, respectively. 
}
\label{fig1}
\end{figure}

However, 
there are only a few studies of systems 
that are composed of nontriangular local structures. 
Since frustration occurs when bonds of antiferromagnetic interaction 
form an odd-number polygon, the next candidate 
is a pentagon. 
Under such circumstances, 
Bi$_2$Fe$_4$O$_9$
in Ref.~\ref{Ressouche_Cairo_exp}
and 
Bi$_4$Fe$_5$O$_{13}$F 
in Ref.~\ref{Abakumov_Cairo_exp} 
were studied; 
there are  
materials of an antiferromagnet on a lattice 
composed of pentagons. 
The lattice is called the Cairo pentagon lattice, 
which is shown in Fig.~\ref{fig1}. 
Although an Fe$^{3+}$ ion in these materials behaves 
as an $S=5/2$ spin, from the theoretical point of view, 
a numerical-diagonalization investigation 
of the $S=1/2$ antiferromagnet on this lattice 
was carried out\cite{Rousochatzakis_Cairo}.   
The behavior under an external magnetic field, however, 
is not understood sufficiently. 

In the present letter, 
we report our study of the magnetization process of the 
$S=1/2$ Heisenberg antiferromagnet on 
a Cairo pentagon lattice 
by the numerical-diagonalization method 
while the ratio of two antiferromagnetic interactions 
in the system is varied.  
Particularly, we focus our attention 
on the behavior around the one-third height of the saturation. 
Extensive studies of this model 
in the entire range 
of the ratio 
will be published elsewhere\cite{Isoda_Cairo_fullpaper}. 
The purpose of the present study 
is to clarify the changes in the behavior 
with the variation in the ratio. 
We report a marked change around a particular ratio, 
which is accompanied 
by a spin-flop phenomenon in a system without spin anisotropy. 


Before we present our new results, 
let us review the features of the Cairo pentagon lattice. 
Vertices of pentagons in the Cairo pentagon lattice 
are divided into two groups: 
one is a vertex with a coordination number $z=3$ 
and the other is a vertex with $z=4$, 
which are hereafter called the $\alpha$ and $\beta$ sites, 
respectively. 
Note here that there are two types of bonds 
connecting the two neighboring vertices: 
$\alpha$-$\alpha$ and  $\alpha$-$\beta$ bonds. 
The unit cell of the lattice shown in Fig.~\ref{fig1} 
include six sites, among which 
there are four $\alpha$ and two $\beta$ sites. 

The Hamiltonian studied in this research is given by 
${\cal H}={\cal H}_0 + {\cal H}_{\rm Zeeman}$, where 
\begin{equation}
{\cal H}_0 = \sum_{\langle i,j\rangle \in {\rm \mbox{$\alpha$-$\alpha$} \ bonds}} 
J
\mbox{\boldmath $S$}_{i}\cdot\mbox{\boldmath $S$}_{j} 
+
\sum_{\langle i,j\rangle \in {\rm \mbox{$\alpha$-$\beta$}  \ bonds}} 
J^{\prime}
\mbox{\boldmath $S$}_{i}\cdot\mbox{\boldmath $S$}_{j} 
. 
\label{H_nofield}
\end{equation}
Let us emphasize here that 
${\cal H}_0$ is isotropic in spin space. 
${\cal H}_{\rm Zeeman} $ is given by 
\begin{equation}
{\cal H}_{\rm Zeeman} = - h \sum_{j} S_{j}^{z} .  
\label{H_zeeman}
\end{equation}
Here, $\mbox{\boldmath $S$}_{i}$ 
denotes the $S=1/2$ spin operator 
at site $i$ shown 
by circles in Fig.~\ref{fig1}. 
Energies are measured in units of $J$ 
for the lattice shown in Fig.~\ref{fig1}(b); 
hereafter, we set $J=1$. 
The number of spin sites is denoted by $N_{\rm s}$. 
We investigate how the magnetization process 
of the above model changes 
when the ratio $\eta = J^{\prime}/J$ is tuned. 

We calculate the lowest energy of ${\cal H}_0$ 
in the subspace 
belonging to 
$\sum _j S_j^z=M$, 
using numerical diagonalizations 
based on the Lanczos algorithm and/or the householder algorithm. 
Here, $M$ takes an integer from zero to the saturation value 
$M_{\rm s}$ ($=S N_{\rm s}$). 
The energy is denoted by $E(N_{\rm s},M)$. 
Lanczos diagonalizations have been carried out 
using the MPI-parallelized code, which was originally 
developed in the study of the Haldane gaps\cite{HN_Terai}. 
The usefulness of our program was confirmed in large-scale 
parallelized calculations\cite{kgm_gap,s1tri_LRO}. 

\begin{figure}[htb]
\begin{center}
\includegraphics[width=7cm]{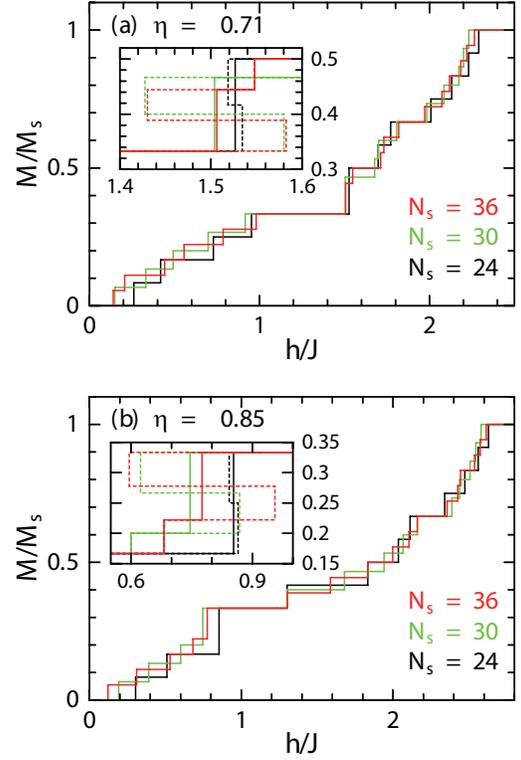}
\end{center}
\caption{(Color) Magnetization process 
of the $S=1/2$ Heisenberg antiferromagnet 
on the Cairo pentagon lattice. 
The results for $\eta=0.71$ and 0.85 are 
presented in (a) and (b), respectively. 
Black, green, and red lines denote the results 
for $N_{\rm s}=24$, 30, and 36, respectively. 
Insets show a zoomed-in view 
at the higher (lower)-field edge 
of the one-third height of the saturation 
when $\eta=0.71(0.85)$, where the dotted lines 
represent the results 
before the Maxwell construction is carried out. 
}
\label{fig2}
\end{figure}

For a finite-size system, 
the magnetization process is determined by 
the magnetization increase from $M$ to $M+1$ at the field 
\begin{equation}
h=E(N_{\rm s},M+1)-E(N_{\rm s},M),
\label{field_at_M}
\end{equation}
under the condition that the lowest-energy state 
with the magnetization $M$ and that with $M+1$ 
become the ground state in specific magnetic fields.  
When the lowest-energy state with the magnetization $M$ 
does not become the ground state in any field, 
the magnetization process around the magnetization $M$ 
is determined by the Maxwell construction. 
In this study, 
we treat the finite-size clusters of $N_{\rm s}=24$, 30, and 36 
shown in Fig.~\ref{fig1}. 
The periodic boundary condition is imposed 
for each cluster. 


Let us, first, examine the magnetization process 
of the present model. 
When we tune $\eta$, we discover that 
a significant change in the behavior 
of the magnetization process 
appears 
at approximately $\eta=0.8$. 
To observe the change, we depict 
our results for $\eta=0.71$ and 0.85 in Fig.~\ref{fig2}. 
In the one-third height of the saturation 
on which we focus our attention, 
one observes the presence of the magnetic plateau 
as a characteristic behavior. 
As a surprising feature, 
there also exists a magnetization jump 
at the {\it higher}-field edge of the plateau 
for $\eta=0.71$,  
while a similar jump appears 
at the {\it lower}-field edge of the plateau 
for $\eta=0.85$. 
Note here that a marked change 
in the magnetization process is induced 
only owing to a small change in $\eta$. 
The presence of the jump was pointed out 
in Ref.~\ref{Rousochatzakis_Cairo}; 
the authors of Ref.~\ref{Rousochatzakis_Cairo} 
speculated that 
the behavior originated from the spin-nematic state 
because of $\Delta M=2$ at the jump. 
In the present study, on the other hand, 
we investigate this behavior of the present model  
from the viewpoint of the spin-flop 
phenomenon\cite{Neel_spin_flop,kohno_aniso2D,sakai_aniso} 
to understand the origin of the marked change.  

\begin{figure}[htb]
\begin{center}
\includegraphics[width=7cm]{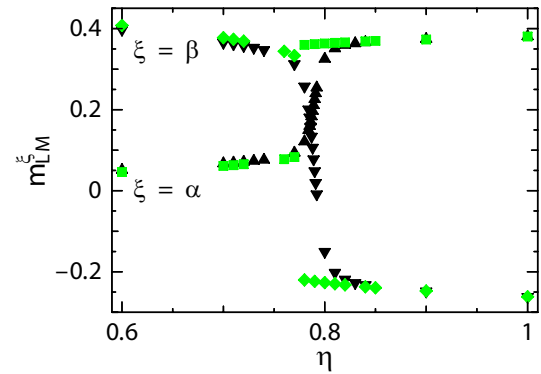}
\end{center}
\caption{(Color) 
Dependence of the average local magnetization 
on the ratio of interaction $\eta$. 
For $N_{\rm s}=24$, 
triangles and inversed triangles 
denote the results of $\xi=\alpha$ and $\beta$, respectively.  
For $N_{\rm s}=30$, 
squares and diamonds 
denote the results and $\xi=\alpha$ and $\beta$, respectively.  
}
\label{fig3}
\end{figure}

For this purpose, we examine the average local magnetization 
at the one-third height of the saturation. 
The average local magnetization is evaluated by 
\begin{equation}
m_{\rm LM}^{\xi} = \frac{1}{N_{\xi}} 
\sum_{j\in \xi} \langle S_j^{z} \rangle , 
\label{ave_local_mag}
\end{equation}
where $\xi$ takes $\alpha$ and $\beta$. 
Here, the symbol $\langle {\cal O} \rangle$ 
denotes the expectation value 
of an operator ${\cal O}$ with respect 
to the lowest-energy state 
within the subspace with a fixed $M$ of interest. 
Note that the average 
over $\xi$ is carried out 
in the case of degenerate ground states 
for some values of $M$, 
where $N_{\xi}$ denotes the number of $\xi$ sites. 
For $M$ with a nondegenerate ground state, 
the results do not change irrespective of 
the presence or absence of the average. 
Our results of the $\eta$ dependence of 
$m_{\rm LM}^{\xi} $ at $M/M_{\rm s}=1/3$ 
are shown in Fig.~\ref{fig3}.  
One clearly observes a large decrease in $m_{\rm LM}^{\beta}$ 
and a large increase in $m_{\rm LM}^{\alpha}$ 
near $\eta=0.78$.  
However, we will publish 
the phase diagram under a nonzero magnetic field 
in a wide range of $\eta$ in Ref.~\ref{Isoda_Cairo_fullpaper}; 
here, we focus our attention 
only on the behavior near $\eta=0.78$ 
and we mention the phases around there with some relationships. 
In the region of $\eta$ larger than this value, 
one observes in Fig.~\ref{fig3} that 
an $\alpha$-site spin becomes an up-spin 
while a $\beta$-site spin becomes a down-spin. 
Note that a ground state under $h=0$ 
is exactly ferrimagnetic in the limiting case 
of $\eta\rightarrow\infty$. 
It is also known that the spontaneous magnetized phase 
of ferrimagnetism is spread for $\eta \simge 1.96$, 
which was reported in Ref.~\ref{Rousochatzakis_Cairo}.  
The present spin arrangement suggests that 
the state is in the ferrimagnetic phase 
{\it under a nonzero magnetic field}. 
In the region of smaller $\eta$, 
on the other hand, 
an $\alpha$-site spin becomes almost vanishing, 
while a $\beta$-site spin becomes an up-spin. 
One finds that the vanishing moments at $\alpha$ sites  
suggest that orthogonal singlet dimers are formed 
at a pair of neighboring $\alpha$ 
spins\cite{comment_ortho_dimer} 
if it is considered that, 
in the case of $\eta=0$, the model is reduced 
to a system composed of isolated single spins at $\beta$ sites 
and isolated dimers of antiferromagnetically interacting spins 
at $\alpha$ sites. 
The present results of a marked change of spin states 
strongly suggest that 
a quantum phase transition occurs at a specific $\eta$ 
and that 
the characteristics of the magnetization plateau 
at $M/M_{\rm s}=1/3$ are different between 
the cases of $\eta=0.71$ and 0.85 observed in Fig.~\ref{fig2}. 
Unfortunately, it is difficult to know 
the boundary point more precisely by extrapolating the results 
for finite-size clusters in the present study.
Furthermore, 
it is also difficult to conclude 
whether the transition is of the second order or first order 
because the change in $m_{\rm LM}^{\xi}$ is continuous 
for $N_{\rm s}=24$, 
while it is discontinuous for $N_{\rm s}=30$. 
Even though such issues remain unresolved  
at the present stage, a quantum phase transition 
certainly occurs near $\eta=0.78$.  

\begin{figure}[htb]
\begin{center}
\includegraphics[width=7cm]{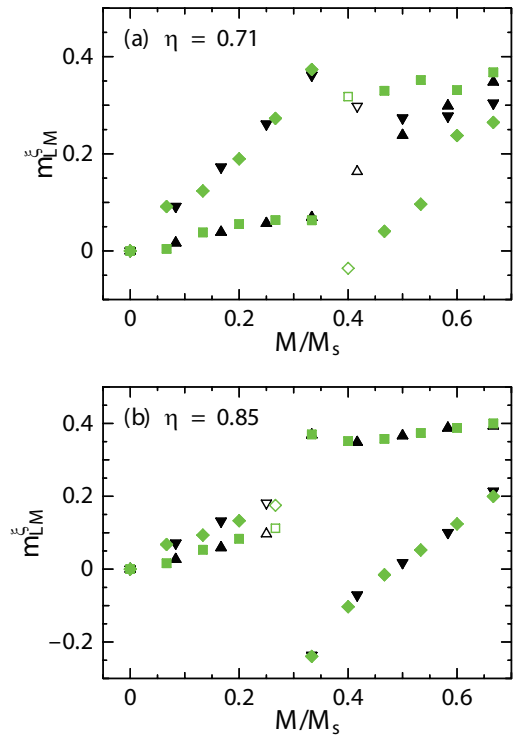}
\end{center}
\caption{(Color) Dependence 
of the average local magnetization 
on the total magnetization. 
The result of $\eta=0.71$ and 0.85 are shown 
in (a) and (b), respectively. 
The shapes of the symbols are the same 
as those in Fig.~\ref{fig3}. 
Closed symbols represent data for the stably realized states, 
while open symbols denote data for the unstable states 
at the magnetization jump. 
}
\label{fig4}
\end{figure}

Next, let us study 
what happens in the appearance of the above magnetization jump 
as a result of the presence of the phase transition. 
For this purpose, we observe the $M/M_{\rm s}$ dependence 
of $m_{\rm LM}^{\xi}$ in the cases of $\eta=0.71$ and 0.85; 
the results are shown in Fig.~\ref{fig4}.    
In the case of $\eta=0.71$, 
the dependences of $m_{\rm LM}^{\xi}$ for $M/M_{\rm s} < 1/3$ 
seem continuous to those of $m_{\rm LM}^{\xi}$ at $M/M_{\rm s} = 1/3$ 
while 
the dependences of $m_{\rm LM}^{\xi}$ for $M/M_{\rm s} > 1/3$ 
seem discontinuous to those of $m_{\rm LM}^{\xi}$ at $M/M_{\rm s} = 1/3$, 
even though we do not take 
the open-symbol data for energetically unrealized states 
into account. 
This discontinuity indicates an abrupt change 
in the spin orientation 
between the cases of $M/M_{\rm s} = 1/3$ and $M/M_{\rm s} > 1/3$, 
which is the origin of the magnetization jump 
as a consequence of the spin-flop phenomenon. 
Note that such phenomena occur 
in spite of the fact that the present system 
does not include any anisotropy in spin space. 
Such spin-flop phenomena in systems 
with the same spin-isotropic nature 
have recently been reported\cite{shuriken_lett} 
in antiferromagnets on a square-kagome 
lattice\cite{squagome_Siddharthan_Georges,
squagome_Tomczak,squagome_Richter}
and a $\sqrt{3}\times\sqrt{3}$-distorted kagome lattice
\cite{Hida_kagome,kgm_distort}. 
In the case of $\eta=0.85$, 
on the other hand, 
the dependences of $m_{\rm LM}^{\xi}$ for $M/M_{\rm s} > 1/3$ 
seem continuous to those of $m_{\rm LM}^{\xi}$ at $M/M_{\rm s} = 1/3$, 
while 
the dependences of $m_{\rm LM}^{\xi}$ for $M/M_{\rm s} < 1/3$ 
seem discontinuous to those of $m_{\rm LM}^{\xi}$ at $M/M_{\rm s} = 1/3$. 
One finds a change on one side of the edges 
where the discontinuous behavior appears in the spin orientation. 
The change also affects the appearance of the magnetization jump, 
as shown in Fig.~\ref{fig2}. 
The key to understanding the dependence of the behavior 
on $\eta$ is 
the difference between $m_{\rm LM}^{\xi}$ of $M/M_{\rm s} > 1/3$  
and that of $M/M_{\rm s} < 1/3$.   
Although $m_{\rm LM}^{\xi}$ on both sides of the vicinity of $M/M_{\rm s} = 1/3$ 
tries to catch up with the quite rapidly changing 
$m_{\rm LM}^{\xi}$ just at $M/M_{\rm s} = 1/3$,  
the catching up cannot necessarily be realized 
on both sides of $M/M_{\rm s} > 1/3$ and $M/M_{\rm s} < 1/3$. 
The present result in the case of $\eta=0.85$ also 
suggests that the spin-flop phenomenon in a spin-isotropic system 
can occur 
not only at the higher-field edge of the one-third height 
of the saturation. 
It should be investigated in future studies 
whether or not the same phenomenon occurs at other heights 
of the saturation. 

Similar magnetization jumps of $\Delta M=2$ are known 
in two finite-size systems without anisotropies in spin space: 
one is the Heisenberg cluster on an 
icosaheron\cite{Schroder_icosahedron} 
and the other is the Heisenberg cluster on a 
dodecahedron\cite{Konstantinidis_icosahedron}. 
Within the condition that one considers regular polyhedra, 
unfortunately, 
it is impossible to increase the number of spins systematically 
to that of an infinite system. 
The jumps in the two finite-size clusters necessarily appear 
between finite-size plateaux, although the widths of the plateaux 
beside the jump in the dodecahedron system are significantly larger 
than the widths of other finite-size plateaux in the same system. 
With this point of view, the situation is contrast to 
the behavior of the present model where 
no indication of the magnetization plateau in the thermodynamic 
limit of $N\rightarrow\infty$ is observed 
on the {\it opposite} side of a clearly existing magnetization plateau 
with respect to the jump. 
The jump of the icosaheron system appears 
only for a large spin amplitude of $S = 4$, while 
no jump is observed in the case of $S < 4$. 
Future studies of the Cairo-pentagon-lattice antiferrmagnet 
composed of larger spins would make it possible to examine 
the relationship between the icosaheron system and the present model. 

\begin{figure}[htb]
\begin{center}
\includegraphics[width=7cm]{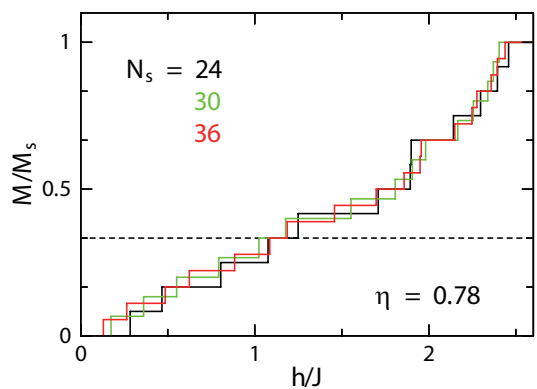}
\end{center}
\caption{(Color)
Magnetization process of the Cairo-pentagon-lattice 
antiferromagnet 
with a ratio of $\eta=0.78$. 
Black, green, and red lines denote the results 
for $N_{\rm s}=24$, 30, and 36, respectively. 
}
\label{fig5}
\end{figure}

Finally, let us observe the magnetization process 
around the boundary of the phase transition 
of $M/M_{\rm s} = 1/3$. 
We depict the result of the magnetization process 
at $\eta=0.78$ in Fig.~\ref{fig5}. 
No jumps are observed. 
One finds that the width of the finite-size step 
at $M = (1/3) M_{\rm s}$ is very small 
in comparison with the widths of those for $\eta=0.71$ 
and 0.85 in Fig.~\ref{fig2}. 
Within the result of $\eta=0.78$, 
the width is also smaller than the widths of steps 
at $M = (1/3) M_{\rm s}+1$ and $M = (1/3) M_{\rm s}-1$. 
The small width suggests the disappearance of the magnetic plateau.  
However, a careful extrapolation toward the thermodynamic limit 
should be performed; this is an open issue in future studies. 


In summary, 
we have investigated the magnetization process 
of the $S=1/2$ Heisenberg antiferromagnet 
on 
the Cairo pentagon lattice 
by the numerical-diagonalization method
with a variation in the ratio of the two antiferromagnetic interactions. 
We have found that a 
quantum phase transition 
occurs at the one-third height of the saturation. 
We have also found that a magnetization jump appears 
due to the spin-flop phenomenon. 
The present model is the first case 
when the spin-flop phenomenon without spin anisotropy 
occurs in a system that does not include 
a local lattice structure of triangles. 
Our unexpected observation is that 
the side where the jump appears depends 
on which side of the transition 
the ratio is located. 
The behavior of the Cairo-pentagon-lattice antiferromagnet 
is quite a characteristic behavior. 
Examinations of the comparison 
between other frustrated systems 
including antiferromagnets on the kagome lattice
\cite{Hida_kagome,kgm_ramp}, 
triangular lattice\cite{Sakai_HN_PRBR}, 
and interpolated case
\cite{HN_Sakai_PSS,HN_Sakai_SCES}, 
as well as on the square-kagome 
lattice\cite{shuriken_lett}
would provide us with various viewpoints 
that will contribute much to our understanding of frustration effects.  

\section*{Acknowledgments}
We wish to thank 
Professor Y.~Hasegawa, 
Dr.~T.~Momoi, 
and 
Dr.~N.~Todoroki 
for fruitful discussions. 
This work was partly supported by Grants-in-Aid 
(Nos. 23340109, 23540388, and 24540348) 
from the Ministry of Education, Culture, Sports, Science 
and Technology of Japan (MEXT). 
Some of the computations were 
performed using facilities of 
the Department of Simulation Science, 
National Institute for Fusion Science; 
Center for Computational Materials Science, 
Institute for Materials Research, Tohoku University; 
the Supercomputer Center, 
Institute for Solid State Physics, The University of Tokyo;  
and Supercomputing Division, 
Information Technology Center, The University of Tokyo. 
This work was partly supported 
by the Strategic Programs for Innovative Research, 
MEXT, 
and the Computational Materials Science Initiative, Japan. 
The authors would like to express their sincere thanks 
to the staff members of the Center for Computational Materials Science 
of the Institute for Materials Research, 
Tohoku University for their continuous support 
of the SR16000 supercomputing facilities. 



\begin{thebibliography}{99} 
\bibitem{Ressouche_Cairo_exp}
\label{Ressouche_Cairo_exp}
E.~Ressouche, V.~Simonet, B.~Canals, M.~Gospodinov, 
and V.~Skumryev, 
Phys. Rev. Lett. \textbf{103}, 267204 (2009).
\bibitem{Abakumov_Cairo_exp}
\label{Abakumov_Cairo_exp}
A.~M.~Abakumov, D.~Batuk, A.~A.~Tsirlin, C.~Prescher, L.~Dubrovinsky, 
D.~V.~Sheptyakov, W.~Schnelle, J.~Hadermann, and G.~V.~Tendeloo1, 
Phys. Rev. Lett. \textbf{87}, 024423 (2013).
\bibitem{Rousochatzakis_Cairo}
\label{Rousochatzakis_Cairo}
I.~Rousochatzakis, A.~M.~L$\ddot{\rm a}$uchli, 
and R.~Moessner, 
Phys. Rev. B \textbf{85}, 104415 (2012).
\bibitem{Isoda_Cairo_fullpaper}
\label{Isoda_Cairo_fullpaper}
M.~Isoda, H.~Nakano, and T.~Sakai, submitted to J.~Phys.~Soc.~Jpn. 
\bibitem{HN_Terai}
\label{HN_Terai}
H.~Nakano and A.~Terai, 
J.~Phys.~Soc.~Jpn. \textbf{78}, 014003 (2009).
\bibitem{kgm_gap}
\label{kgm_gap}
H.~Nakano and T.~Sakai, 
J.~Phys.~Soc.~Jpn. \textbf{80}, 053704 (2011).
\bibitem{s1tri_LRO}
\label{s1tri_LRO}
H.~Nakano and T.~Sakai, 
J.~Phys.~Soc.~Jpn. \textbf{82}, 043715 (2013).
\bibitem{Neel_spin_flop}
L.~N$\acute{\rm e}$el, Ann. Phys. ~Paris \textbf{5}, 232 (1936).
\bibitem{kohno_aniso2D}
M.~Kohno and M.~Takahashi, 
Phys.~Rev.~B \textbf{56}, 3212 (1997). 
\bibitem{sakai_aniso}
T.~Sakai and M.~Takahashi, 
Phys.~Rev.~B \textbf{60}, 7295 (1999). 
\bibitem{comment_ortho_dimer}
This state is also similar to the inverse-T state 
in the square-kagome-lattice antiferromagnet 
in Ref.~\ref{shuriken_lett}. 
\bibitem{shuriken_lett}
\label{shuriken_lett}
H.~Nakano and T.~Sakai, 
J.~Phys.~Soc.~Jpn. \textbf{82}, 083709 (2013).
\bibitem{squagome_Siddharthan_Georges}
\label{squagome_Siddharthan_Georges}
R. Siddharthan and A. Georges: 
Phys. Rev. B \textbf{65} (2001) 014417. 
\bibitem{squagome_Tomczak}
\label{squagome_Tomczak}
P. Tomczak and J. Richter, 
J. Phys. A: Math. Gen. \textbf{36}, 5399 (2003). 
\bibitem{squagome_Richter}
\label{squagome_Richter}
J. Richter, J. Schulenburg, P. Tomczak, and D. 
Schmalfu{\ss}, 
Condens. Matter Phys. \textbf{12}, 507 (2009).
\bibitem{kgm_distort}
\label{kgm_distort}
H.~Nakano, T.~Sakai, and Y.~Hasegawa, 
submitted to J.~Phys.~Soc.~Jpn. 
\bibitem{Hida_kagome}
\label{Hida_kagome}
K.~Hida, 
J.~Phys.~Soc.~Jpn. \textbf{70}, 3673 (2001).
\bibitem{Schroder_icosahedron}
C.~Schr$\ddot{\rm o}$der, H.~J.~Schmidt, J.~Schnack, and M.~Luban,
Phys.~Rev.~Lett. \textbf{94}, 207203 (2005). 
\bibitem{Konstantinidis_icosahedron}
N.~P.~Konstantinidis, 
Phys.~Rev.~B \textbf{72}, 064453 (2005). 
\bibitem{kgm_ramp}
\label{kgm_ramp}
H.~Nakano and T.~Sakai, 
J.~Phys.~Soc.~Jpn. \textbf{79}, 053707 (2010).
\bibitem{Sakai_HN_PRBR}
\label{Sakai_HN_PRBR}
T.~Sakai and H.~Nakano, 
Phys.~Rev.~B \textbf{83}, 100405(R) (2011). 
\bibitem{HN_Sakai_PSS}
\label{HN_Sakai_PSS}
H.~Nakano and T.~Sakai, 
Phys. Status Solidi B \textbf{250}, 579 (2013). 
\bibitem{HN_Sakai_SCES}
\label{HN_Sakai_SCES}
H.~Nakano and T.~Sakai, 
to be published in JPS Conf. Proc. 
%
\end{thebibliography}
\end{document}